Tidal Heating of Extra-Solar Planets

by

Brian Jackson, Richard Greenberg, & Rory Barnes

Lunar and Planetary Laboratory
University of Arizona
1629 E University Blvd, Tucson AZ 85721-0092

Running Title: Tidal Heating

Abstract:

Extra-solar planets close to their host stars have likely undergone significant tidal evolution since the time of their formation. Tides probably dominated their orbital evolution once the dust and gas had cleared away, and as the orbits evolved there was substantial tidal heating within the planets. The tidal heating history of each planet may have contributed significantly to the thermal budget that governed the planet's physical properties, including its radius, which in many cases may be measured by observing transit events. Typically, tidal heating increases as a planet moves inward toward its star and then decreases as its orbit circularizes. Here we compute the plausible heating histories for several planets with measured radii, using the same tidal parameters for the star and planet that had been shown to reconcile the eccentricity distribution of close-in planets with other extra-solar planets. Several planets are discussed, including for example HD 209458 b, which may have undergone substantial tidal heating during the past billion years, perhaps enough to explain its large measured radius. Our models also show that GJ 876 d may have experienced tremendous heating and is probably not a solid, rocky planet. Theoretical models should include the role of tidal heating, which is large, but time-varying.

Subject Headings: celestial mechanics

1. Introduction

A substantial fraction of known extra-solar planets are so close to their host stars that they must have undergone significant amounts of tidally driven evolution. Changes in their orbits include the linked evolution of their eccentricities $e$ and semi-major axes $a$. It has generally been assumed that the eccentricities of the close-in planets (those with $e < 0.2$ AU) began larger and have been reduced by tidal damping (*e.g.* Rasio *et al.* 1996, Marcy *et al.* 2005, Jackson *et al.* 2008). Recently we confirmed that the current distribution of eccentricities could have evolved from a distribution identical to that of the farther-out planets, and that this result could be achieved using a reasonable and consistent set of tidal parameter values (Jackson *et al.* 2008).

Significant contributions to a planet's orbital evolution can come from tides raised on a star by a planet and from tides raised on the planet by the star (Jackson *et al.* 2008). During the course of the tidal evolution, tidal distortion of the figure of the planet can result in substantial amounts of internal heating at the expense of orbital energy. The heating rate as a function of time is coupled to the evolution of the orbit. Compared with their current orbits, many close-in planets probably were farther from their host star (larger semi-major axes) at the time that the planetary system formation had ceased and the gaseous nebula dissipated. In a typical case, tidal heating might have begun modest, but then increased as tides reduced the semi-major axis $a$. As the tides became stronger, they would circularize the orbit, which in turn would shut down the tidal heating mechanism.

Thus two competing effects are in play: the reduction of $a$ which increases tidal heating, and the damping of $e$ which decreases the heating. The relative strength and timing of these two effects would determine a planet's history, typically with a gradual increase in the heating rate followed by a decrease.

The thermal history of a planet is critical to determining its physical properties. For example, models of extra-solar planets have considered the effects of heating on their radii, which can be measured directly by transit observations. Heat sources that have been considered in these models include the energy of planetary accretion and radiation from the star, as well as tidal heating (Bodenheimer *et al.* 2003, Mardling 2007, Winn & Holman 2007). In many cases the theoretical predictions match the observations reasonably well. However, there are notable exceptions. HD 209458 b has been observed by Knutson *et al.* (2007) to have a radius of 1.32 Jupiter radii ($R_{Jup}$), which is about 10-20% larger than predicted by theoretical modeling (Guillot 2005, see also Burrows *et al.* 2007). HAT-P-1 b is also larger than predicted by theory (Bakos *et al.* 2007a). On the other hand, HAT-P-2 b is observed (Bakos *et al.* 2007b) to be about 12% *smaller* than the radius predicted by theory (Fortney *et al.* 2007).

Theoretical models must make multiple assumptions about the behavior of gases at high pressures, atmospheric heat flow, and radiative cooling, among others. Moreover, theoretical models to date have not taken into account the history of tidal heating for close-in planets. Of course, those are the planets most likely to have radii measurable by transits. As a first cut at addressing this issue, we here present the tidal heating histories that would accompany the orbital evolution as computed by Jackson *et al.* (2008).

2. Method

The computation of the tidal evolution was described by Jackson *et al.* (2008). The method is based on the tidal evolution equations for *da/dt* and *de/dt* assembled by Kaula (1960) and Goldreich & Soter (1966). These equations had been applied to extra-solar planets previously, although certain terms were often neglected. For example, consideration of the changes in orbital eccentricity often disregarded the term due to tides on the star, while changes in semi-major axes were estimated ignoring tides raised on the planet. Moreover, changes in *e* were often represented by a damping timescale, which implicitly neglected the coupling of the evolution of *e* and a. Jackson *et al.* considered tides raised on both the planet and the star, as well as the strong coupling between evolution of the orbital elements. By integrating the orbits of close-in planets (*i.e.* the so-called "hot Jupiters") back in time to the end of their formative period, the distribution of their *e* values was found to match that of the planets with larger *a* values. The best fit was obtained with values of $Q'_p$ of $10^{6.5}$ for all the planets and $10^{5.5}$ for all the stars, although the value of $Q_*$ was relatively unconstrained. Here the tidal dissipation parameter $Q'_p$ includes a factor that accounts for the deviation of the tidal Love number *k* from a nominal value of 3/2. It is likely that individual planets and stars have different dissipation parameters, but the values found by Jackson *et al.* (2008) are consistent with previous estimates (Yoder & Peale 1981, Lin *et al.* 1996) and provide a basis for estimating the potential significance of tidal heating in governing the physical properties of the planets.

The theory applied by Jackson *et al.* will certainly need to be updated as understanding of the response of stars and planets to a time-varying tidal potential improves. For example, Jackson *et al.* (2008) showed that past eccentricities may have been quite large, but there is considerable uncertainty regarding rates of tidal evolution under such conditions. Understanding of the frequency dependence of $Q_p$ will allow formulation of governing equations that have more credibility for large values of *e* (*e.g.,* Mardling & Lin 2002) or for cases with a comparable rotation period of the star and orbital period of the planet (Dobbs-Dixon *et al.* 2007). Considerable progress is being made in this area (Ogilvie & Lin 2004, 2007). However, given current understanding, the formulation used by Jackson *et al.* (2008) is reasonable and can be readily updated, pending improved tidal models.

The dissipation of energy within a planet due to tides comes directly from the planet's orbit (except very early-on while the planetary spin is quickly reduced). The orbital energy depends on only one orbital element, the semi-major axis. Thus the term in the equation for *da/dt* that corresponds to tides raised on the planet by the star also gives the heating rate. In this way, the heating rates can be extracted from the same numerical integrations that were done for the tidal evolution of orbits by Jackson *et al.* (2008). In addition, we have considered for various planets of interest the range of heating histories, given the range of uncertainties in their current orbital elements. For example, planets with nominal reported eccentricities of zero were not considered by Jackson *et al.* (2008). If *e* were really zero, there would be no tidal heating either now or in the past. Here we consider the possible heating, even for those planets, using the full range of *e* and *a* values consistent with observational uncertainty. The planets that we discuss here are of particular interest because in each case there is some basis for comparison of thermal models with an observationally measured radius.

3. Tidal Heating Histories

The tidal heating rate $H$ can be expressed as

$$H = (\tfrac{63}{4}) \frac{(GM_*)^{3/2} M_* R_p^5}{Q'_p} a^{-15/2} e^2 \quad (1)$$

where $G$ is the gravitational constant, $M_*$ is the stellar mass, $R_p$ the planetary radius, and $Q'_p = 3Q_p/2k$. $Q_p$ and $k$ are the tidal dissipation parameter and Love number, respectively.

Tidal evolution of the orbital elements reduces both *a* and *e*, with competing effects on the heating rate: Reducing *a* increases the heating, but reducing *e* decreases it. The evolutionary histories computed by Jackson *et al.* (2008) are such that the change in *a* usually dominates earlier on, while the damping of *e* dominates later. Thus a typical heating history involves an increase to a maximum rate, followed by a decrease to the current rate. In the next section, we consider the variation in heating rate for several planets of interest. The orbital and physical parameters we adopted for our modeling are listed in Table 1, in which $M_p$ is a planet's mass and $R_*$ is the stellar radius. Minimum (min), nominal (nom) and maximum (max) values of orbital parameters that are allowed by observation are listed there as well.

3.1. Planets with published eccentricities

3.1.1. HD 209458 b
Fig. 1 shows the time history of the tidal heating rate. The solid curve is based on the nominal current $e = 0.014$ and $a = 0.0473$ AU (Laughlin *et al.* 2005). The dashed line is based on the maximum current heating rate consistent with the observational constraints on the orbital elements, *i.e.* with the maximum plausible current value of *e* (0.042) and the minimum plausible value of *a* (0.0459 AU). The integrations go back 15 Gyr, although the nominal age of the system (shown by the vertical line) is 2.5 Gyr (Takeda *et al.* 2005). Burrows *et al.* (2007) suggest that a heating rate of about $4 \times 10^{19}$ W would be required to maintain the observed planetary radius. Therefore, even with the largest possible current *e* value, the current tidal heating rate is too small to resolve the discrepancy between the large observed planetary radius and theoretical models. However, the history plotted in Fig. 1 shows that the required heating rate was maintained for about a billion years after the system formed, and dropped off only about a billion years ago. If the lag in the response of the planet to the heating rate were on the order of a billion years, it may explain the observed large radius. Such a lag seems reasonable based on the long duration of the influence of heat of formation on the planet's radius in the modeling by Burrows *et al.* (2007).

The evolution shown in Fig. 1 is an example of the most general case of a rising heating rate followed by a decrease. In this case, it is likely that the system was near or at the peak at the end of its formative period. If the current *e* value is less than the nominal value, then the heating rate has probably been decreasing throughout the lifetime of the planet. Note too that for this planet, observations cannot rule out a circular orbit. If that is the case, tidal dissipation could have been negligible. Nevertheless, the best fit case, with a current $e = 0.014$ or larger, corresponds to a heating history that may help explain the large observed radius of this planet.

We note that some authors have estimated that the tidal circularization timescale for HD 209458 b was so short that a mechanism was needed to explain the planet's substantial current eccentricity. For example, Mardling (2007) proposed that an additional planet is needed to maintain the eccentricity. However, transit timing observations rule out the existence of another planet with a period less than 15 days (Miller-Ricci *et al.* 2008). In fact, the issue may be moot, because our calculations show that tides have taken billions of years to circularize HD 209458 b's orbit (Jackson et al. 2008). As discussed in Jackson et al. (2008), the short circularization timescales estimated by previous workers were based on the exponential solution of the equation for tidal damping (*de/dt*), which ignored concurrent and codependent changes in semi-major axis. Because tides reduce both *a* and *e* together, ignoring changes in *a* significantly underestimated the time required to circularize orbits. Using the coupled equations for tidal evolution, we find it is perfectly reasonable that the eccentricity remains fairly large. Thus, tidal heating has likely been large during the past billion years, perhaps explaining the planet's surprisingly large radius.

3.1.2. HAT-P-1 b

Like HD 209458 b, this planet's observed radius of 1.36 $R_{Jup}$ (Bakos *et al.* 2007a) is larger than expected from theoretical modeling that did not include tidal heating (Fortney *et al.* 2007). Fig. 2 shows the tidal heating history for this planet. Similar to HD 209458 b, the heating rate ~ 1 Gyr ago was substantially higher than the present tidal heating, based on either the nominal best fit *e* value (solid curve) or the maximum current *e* value (dashed line). In this case the planetary system is probably much older, so the history extends further back in time, and thus includes a 7 Gyr period of increasing tidal heating (due to the decrease in semi-major axis), followed by a decrease in heating (as the orbit circularizes). For both HD 209458 b and HAT-P-1 b, the substantial heating rate ~ 3-4 x $10^{19}$ W about 1 Gyr ago may help account for the discrepancy between the large observed planetary radii and the predictions of physical modeling.

### 3.1.3. GJ 436 b

This planet has a measured radius consistent with theoretical models, independent of tidal heating (Gillon *et al.* 2007). The tidal heating history shown in Fig. 3 is consistent with that result. Compared with the previous two cases, the maximum heating rate was two orders of magnitude less. This lower rate of dissipation may be partially offset by the fact that the duration of the maximum heating was several billion years, *i.e.* the peak in Fig. 3 is much broader than in Fig. 1 or 2. However, the total tidal heat is still much smaller than the other cases, and most of it was so long ago that its residual effects are probably negligible, which may explain why the measured radius fits the tide-free physical model.

### 3.1.4. TrES-1

This planet may have experienced more tidal heating than any of the previous cases as shown in Fig. 4. Based on that result, and by analogy with HD 209458 b and HAT-P-1 b, one might have expected the measured radius to be larger than predicted by previous physical modeling. However, Winn *et al.* (2007) report that the measured radius does fit theory. It seems that the expected large amount of tidal dissipation did not affect the radius in this case, a surprising result that calls for an explanation. One possibility is that we have overestimated the tidal heating; another is that the theoretical modeling may need adjustment so as to accommodate more heat without increasing the inferred radius.

### 3.1.5. HAT-P-2 b

In this case also there has been a substantial amount of tidal heating (Fig. 5). The current heating rate is similar to the maximum rate attained by HD 209458 b and HAT-P-1 b, so again we might expect a larger radius than predicted by theory that ignored tidal heating. In this case, however, the measured radius is actually smaller than predicted (Bakos *et al.* 2007b). Thus there is a discrepancy between theory and observation even if tidal heating is neglected. Correction to the theoretical modeling seems to be necessary, and the correction would need to be in the same sense as that for TrES-1 (*i.e.*, a smaller radius for a given amount of heating). The fact that there is likely a high rate of tidal dissipation makes the problem even worse.

On the other hand, a key factor in the reconciliation may be that, while the current tidal heating rate is high and increasing, in the recent past the heating rate was much lower. HAT-P-2 b is still on the increasing part of the heating curve, which is unusual among planets considered here, most of which have passed their peaks. The fact that the heating rate was several times smaller a billion years ago than it is now may help explain the small radius.

Compare that result with what we found above for the planets where the observed radii are larger than expected (HD 209458 b and HAT-P-1 b). In those cases, the fact that the heating rate was several times *larger* a billion years ago than it is now may help explain the *large* radius.

In all of these cases it seems that the heating rate in the past (~1 Gyr ago) may have been the crucial factor in determining the current radius.

3.1.6. HD 149026 b

For this planet, the maximum $e$ allowed by observation is 0.02, although the nominal adopted $e$ value is zero. Even assuming a current orbit that would allow maximum heating, (dashed line in Fig. 2) given observational constraints on the current orbit, tidal heating has been only about $2 \times 10^{17}$ W for most of the 8 Gyr age of the system, and has dropped by about 30% during the past 1 Gyr as the eccentricity has damped down. Similarly to GJ 436 b, tidal heating is probably not a factor in determining its radius. In fact, transit observations show HD 149026 b has the smallest radius measured for any extra-solar planet, 0.726 $R_{Jup}$ (Charbonneau *et al* 2006). Interior models require a large core, with a mass ~80 Earth masses, to be consistent with this small observed radius (Sato *et al.* 2005). Conceivably, such a core would have a lower $Q_p$ than we have assumed as the bulk value for this planet, because rocky bodies generally have $Q_p$ ~ 100. In that case the tidal heating may have been great enough to have been a factor in the planet's geophysical history. However, if additional heat were incorporated into the theoretical modeling, it might tend to increase the model radius, perhaps requiring an even larger core to match the observed radius. In any case the theoretical models need to take into account the possible addition of considerable tidal heating.

3.2. Planets with undetermined eccentricities

For most extra-solar planets whose eccentricities have not yet been measurable, the value is customarily tabulated as zero (Butler *et al.* 2006). For at least nine such planets, radii have been measured and can be compared with theoretical predictions. In most cases the models fit the observations, but in four of the nine cases, the theory predicted radii smaller than observed. Next we consider for each of these planets whether tidal heating, which has not yet been incorporated into the modeling, might potentially play a role. Even for the planets with tabulated $e$ values of zero, the true values may be as large as 0.03 (Butler *et al.* 2006). Thus, for each of the nine planets, we consider the implications that would follow if the current $e$ actually has a value of 0.001, 0.01, and 0.03, in order to sample the range of possible values. The results are shown in Figs. 3 and 4. For each planet, the results for current of 0.01 are shown with a solid line, and the results for the smaller and larger values (and thus smaller and larger current heating rates, respectively) are shown with dashed lines.

The following planets have larger radii than had been predicted by models: TrES-2 (Sozzetti *et al.* 2007), WASP-2 b (Sozzetti *et al.* 2007), OGLE-TR-56 b (Pont *et al.* 2007), and XO-1 b (Holman *et al.* 2006). As shown in Fig. 3, the first three may have had tidal heating well in excess of $10^{19}$ W within the past 1 Gyr and probably lasting for at least ~ 1 Gyr, assuming their current $e$ is actually $\geq 0.01$ (and probably even if their current $e$ is as small as a few times $10^{-3}$). Thus tidal heating needs to be included in the physical modeling and may help reconcile the difference between theory and observation. For XO-1 b, tides are unlikely to have played a significant role unless the current $e$ is larger than 0.03. Thus, for this case, the reconciliation probably requires some other correction to the theoretical models, so as to give a smaller radius, which is the same trend suggested for several other cases discussed above.

The five cases that have radii consistent with theoretical models (Burrows *et al.* 2007, Winn *et al.* 2007) are OGLE-TR-10 b, -111 b, -113 b, and -132 b and HD 189733 b. As shown in Fig. 4, in all cases except 111 b, heating rates may have been greater than $10^{19}$ W during the past Gyr even if the current *e* values are only 0.01. What is more, Fig. 4 shows that two planets (OGLE-TR-113 b and HD 189733 b) would have reached $10^{20}$ W even if their current *e* is only 0.001. (OGLE-TR-111 b would require the current *e* to be > 0.03 to have peaked at $10^{19}$ W, but even in that case the burst of heat was several billion years ago.) For several of these planets, there has likely been enough heating to be a factor in controlling the physical properties. Thus the fact that the measured radii fit the models suggests that either the current *e* values are smaller than the values considered here, or, once again, the theoretical models need to be revisited so as to keep the same radii while accommodating the additional heating due to tides.

3.3. Terrestrial-scale planets

Among known planets with masses less than about ten times that of the Earth, tidal heating could have played some role in the geophysical evolution. Here we assume that the planetary $Q_p$ is 100 and the Love number *k* is 0.3, reasonable choices for a rocky planet (Lambeck 1979, Dickey *et al.* 1994, Mardling & Lin 2002, Barnes *et al.* 2007).

For GJ 876 d, the nominal *e* is zero, so following the same procedure as in the previous section for such cases we consider the heating history under the assumption that the current *e* value is actually 0.001, 0.01, or 0.03, as shown by the curves so labeled in Fig. 5. We assume $R_p$ ~ 10,200 km, based on the mass/radius relationship for terrestrial planets given by Sotin *et al.* (2007). In addition, we consider the implications of the upper limit to *e* of 0.28, reported by Rivera *et al.* (2005). The implied heating history in that case (as modeled by Eq. 1) is given by the curve labeled "0.28" in Fig. 5. In each of these four cases, as shown in Fig. 5, the heating rate is well over $10^{19}$ W for tens of millions of years, and peaks at ~$10^{20}$ W. The main difference among the four cases, depending on the assumed current *e* value, is the timing of the peak, although it is within the past ~30 Myr in all four cases.

To put these numbers in context, consider the geophysical modeling of this planet by Valencia *et al.* (2007a) who report that $7 \times 10^{17}$ W would be adequate to induce substantial internal melting of the mantle. According to that result, for GJ 876 d tidal heating (Fig. 5) has likely been 2 orders of magnitude greater than needed for melting. At the surface, in a steady state the tidal heating would correspond to a heat flux of ~$10^{4-5}$ W/m$^2$. For comparison the surface heat flux for spectacularly active Io is ~3 W/m$^2$ (McEwen *et al.* 1992) and is largely due to tides. For the Earth the flux is ~0.08 W/m$^2$ (Davies 1999), which is largely due to radiogenic heat. Valencia *et al.* (2007b) suggested that radiogenic heating of GJ 876 d might have been adequate to initiate plate tectonics, but our results indicate that tidal heating may have been a major contributor to the geological and geophysical character of the planet. Tidal heat has provided an important component of the heat budget for this planet, perhaps the dominant component during at least the past ~$10^8$ yr. The tidal heating rate would be so large, in fact, that GJ 876 d is unlikely to be a solid, rocky body.

For the nominal current orbit of Gl 581 c (Butler *et al.* 2006), the tidal heating is about $10^{16}$ W assuming the best-fit current *e* and *a*. It has been nearly constant over the past billion years, and was only slightly larger early in its history 9 Gyr ago. Considering the range of uncertainty of the current orbit, the values could have been a few times higher or lower, but still ~ $10^{16}$ W, far less than GJ 876 d. Nevertheless, this rate could be geophysically important. Assuming $R_p$ ~ 10,000 km (again scaling from the mass), the tidal contribution to the geophysical heat flux would be ~ 10 W/m$^2$ at the surface, more than twice that of Io and 100 times the heat flux at the surface of the Earth.

HD 69830 b is greater than 10 Earth masses, so it is more likely a Uranus/Neptune class planet than a terrestrial one. Nevertheless, Valencia *et al.* (2007a) considered the possibility that it is terrestrial in character, so we consider the implications for tidal heating if its $Q_p$ were the value expected for a rocky planet. For the best-fit current *e* value, we find that tidal dissipation may generate ~$10^{17}$ W, or up to a few times more given the most optimistic current orbital parameters. The heating rate has been fairly constant over the past billion years, but it had decreased by a factor of a few since the planet's formation ~10 Gyr earlier (Lovis *et al.* 2006). Assuming a radius of ~12,000 km (based on its mass), the surface heat flux over the past 1 Gyr would be 55 W/m$^2$ (20 times Io's). If HD 69830 b is a terrestrial planet, tidal heat must have been a major factor in its geophysics throughout its history.

Note that in these calculations we have ignored the effect of interactions between these planets and other known (and probably still unknown) planets in their systems. Secular interactions and orbital resonances may well have affected the orbital evolution in ways that modified the tidal heating histories. These effects would be in addition to the other uncertainties and assumptions intrinsic to the results presented here. Nevertheless, the point of our results is that tidal dissipation is likely to have been a significant factor in the geophysical evolution of extra-solar terrestrial-type planets.

4. Discussion

The calculations here suggest that tidal heating may well have played an important role in the evolution of the physical properties of many extra-solar planets, the physical property for which we have the best constraints at present being the planetary radius. We caution that the specific calculations displayed here depend on numerous assumptions and several uncertain parameters. The heating rates correspond to the orbital evolution trajectories computed by Jackson *et al.* (2008), and the caveats are discussed in detail there. In particular, the exact tidal histories presented here depend on the choice of $Q_*$. However most of the tidal histories presented here are not very sensitive to the choice of $Q_*$ since tides raised on the planet usually dominate the evolution. HAT-P-2 b is an important exception. Owing to the planet's unusually large mass (~ 9 M$_{Jup}$), tides raised on the star may dominate the tidal evolution. As a result, the rate of tidal evolution depends sensitively on $Q_*$, *e.g.,* larger $Q_*$ results in slower tidal evolution. If tidal evolution were slower for this planet, it would mean that the past heating rate was closer to the current, large heating rate since the orbital parameters would not have changed much.

In any case, it is quite likely that the actual thermal history of any particular planet was different to some degree from what we show here. However, the unavoidable point is that tidal heating may be significant and should be considered as a factor in theoretical modeling of physical properties.

As shown in Section 3, in most cases where the measured radius is greater than the theoretically predicted value (HD 209458 b, HAT-P-1 b, TrES-2, WASP-2 b, and OGLE-TR-56 b), the tidal heating has been significant and may thus resolve (or at least contribute to resolving) the discrepancy, once it is incorporated into the physical models. The greatest heating was typically ~ 1 Gyr ago, so it may be that current planetary radii reflect the heating rate at that time in the past.

In only one case where the measured radius is greater than theoretically predicted (XO-1 b) is the tidal heating too small to be a significant factor. Some other correction is probably needed to bring the model into agreement with the observation.

Among cases where the theoretical models have seemed to be in good agreement with measured radii, two have experienced negligible tidal heating (GJ 436 b and OGLE-TR-111 b), so the agreement is preserved even when tides are taken into account. However, five cases may have undergone considerable tidal heating (TrES-1, HD 189733 b, OGLE-TR-10 b, -113 b, and -132 b). Physical models will need to incorporate this heat source, which is likely to increase the radius predicted by the models. If the change is significant, other modifications will be required before the theory can be considered in agreement with observations.

In two of the cases considered here, the theoretical models predicted radii larger than actually measured. For HD 149026 b, tidal heating is probably not a factor. For HAT-P-2 b, the tidal heating exacerbates the discrepancy between theory and observation. However, the problem is less severe if (as suggested by the other cases above) the current radius reflects the heating rate ~1 Gyr ago, when the heating rate (in this case) was a few times less than at present.

Our study also suggests that the current state of physical modeling often gives radii that are too large for a given assumed amount of heating (or equivalently underestimating the amount of heat needed to yield a given radius). Tidal heat may resolve the discrepancy between theory and observations for most of the cases where measured radii were larger than expected, but it may make things worse in most of the cases where measured radii seemed to fit the current models. In two cases, whether tidal heating is significant or not, the observed radii are smaller than predicted by the models. In summary, some discrepancies may be resolved by taking into account tidal heating, some remain even when tidal heating is taken into account, and some are exacerbated by tidal heating. Even when tidal heating is included, theoretical models will generally need to be adjusted and improved so as to yield smaller radii for a given amount of internal heat, if they are to agree with measured values. Certainly, theoretical studies of the evolution of the physical properties of planets need to account for tides as a significant source of heat, and one that varies over time. Unlike heat of accretion, tidal heating varies over time, often reaching its maximum considerably later in the life of the planet.

Among terrestrial-scale planets, we find that tidal heating may have dominated the geological and geophysical evolution of the planets and control their current character. The tidal heating rate for GJ 876 d may be orders of magnitude greater than the magnitude considered by Valencia *et al.* to be geophysically significant. For Gl 581 c tidal heating may yield a surface flux about three times greater than Io's, suggesting the possibility of major geological activity.

The surface flux of tidal heat on HD 69830 b would be yet an order of magnitude larger if it is a rocky planet. These heating rates are so large (especially for GJ 876 d) that the extensive melting implied may not be consistent with the tidal dissipation parameters that we have assumed. Some melting might increase the rate of tidal heating as the tidal amplitude increases, but deep global melting would increase $Q_p$ and limit the heating to rates lower than what we have calculated. Another caveat is that the masses of these planets are minimum masses from radial velocity studies, so they may not be rocky, terrestrial scale bodies after all. Finally we emphasize that we have ignored the effects of mutual interactions among planets, which may affect orbital and tidal evolution in interesting ways.

Our results demonstrate the importance of using the coupled equations of tidal evolution of $e$ and $a$. Some previous considerations of tidal evolution considered only the equation for $de/dt$, with semi-major axis held constant, which results in a simple exponential solution. The exponential damping (or "circularization timescale") found in that way can grossly underestimate the actual time that it takes to decrease the orbital eccentricity, as demonstrated by Jackson et al. (2008). Here we have seen that the slower circularization has likely resulted in significant heating rates at present and in the relatively recent past.

These results show that tidal heat may be a major factor in determining the character of extra-solar planets. The state of our understanding of tidal processes and of the relevant parameters for the planets is such that the specific results obtained here should be regarded as tentative at best. The tidal dissipation processes and magnitude for these planets remains largely a matter of speculation. The extent to which a given amount of heat affects the physical state of the planet must depend on where the heat is dissipated. Heating near the surface may have less effect than if it is deep in the interior. Despite these uncertainties, the results presented here do demonstrate that tidal heat must be considered in theoretical modeling. The heating histories calculated here give a basis for a first cut at integrating tidal heat into physical models of extra-solar planets.

Acknowledgments: We thank Adam Burrows for a helpful discussion of the context of this work. We would also like to thank the referee for useful comments. This project is supported by NASA's Planetary Geology and Geophysics program.

| Planet Names | M* (M$_{sol}$) | R* (R$_{sol}$) | M$_p$ (M$_{Jup}$) | R$_p$ (R$_{Jup}$) | a (AU) min | a (AU) nom | a (AU) max | e min | e nom | e max | Age (Gyr) | References M*, R*, M$_p$, R$_p$, a, e, Age |
|---|---|---|---|---|---|---|---|---|---|---|---|---|
| HD 209458 b | 1.140 | 1.130 | 0.64 | 1.32 | 0.0459 | 0.0473 | 0.0487 | 0 | 0.014 | 0.042 | 2.44 | 1, 1, 2, 2, 3, 3, 4 |
| HAT-P-1 b | 1.120 | 1.150 | 0.53 | 1.36 | 0.0536 | 0.0551 | 0.0566 | 0 | 0.09 | 0.11 | 3.6 | 5 (all) |
| GJ 436 b | 0.440 | 0.440 | 0.0706 | 0.3525 | 0.0276 | 0.025b5 | 0.0293 | 0.14 | 0.16 | 0.18 | 9.23[a] | 6 (all) |
| TrES-1 | 0.880 | 0.810 | 0.75 | 1.08 | 0.0386 | 0.0393 | 0.04 | 0.039 | 0.135 | 0.231 | 2.5 | 8, 8, 8, 8, 8, 8, 9 |
| HAT-P-2 b | 1.350 | 1.800 | 9.05 | 0.982 | 0.0673 | 0.0685 | 0.0697 | 0.495 | 0.507 | 0.519 | 2.7 | 10 (all) |
| HD 149026 b | 1.300 | 1.450 | 0.36 | 0.73 | 0.042 | 0.043 | 0.044 | 0 | 0 | 0.02 | 2.0 | 11 (all) |
| TrES-2 | 1.080 | 1.000 | 1.25 | 1.24 | 0.0362 | 0.0367 | 0.0379 | 0.001 | 0.01 | 0.03 | … | 12, 12, 12, 12, 12, …, … |
| WASP-2 b | 0.790 | 0.810 | 0.88 | 1.04 | 0.0296 | 0.0307 | 0.0318 | 0.001 | 0.01 | 0.03 | … | 13, 13, 13, 13, 14, …, … |
| OGLE-TR-56 b | 1.170 | 1.320 | 1.29 | 1.3 | 0.0221 | 0.0225 | 0.0229 | 0.001 | 0.01 | 0.03 | 2.5 | 15, 15, 15, 15, 15, …, 9 |
| XO-1 b | 1.000 | 0.930 | 0.9 | 1.18 | 0.0483 | 0.0488 | 0.0493 | 0.001 | 0.01 | 0.03 | 4.6 | 16, 16, 16, 16, 17, …, 17 |
| OGLE-TR-10 b | 1.020 | 1.160 | 0.63 | 1.27 | 0.0423 | 0.043 | 0.0436 | 0.001 | 0.01 | 0.03 | 2.0 | 15, 15, 15, 15, 15, …, 9 |
| OGLE-TR-111b | 0.810 | 0.830 | 0.52 | 1.07 | 0.046 | 0.047 | 0.048 | 0.001 | 0.01 | 0.03 | 5.55 | 18, 18, 18, 18, 18, …, 23 |
| OGLE-TR-113 b | 0.780 | 0.770 | 1.32 | 1.09 | 0.0227 | 0.0229 | 0.0231 | 0.001 | 0.01 | 0.03 | 5.35 | 19, 19, 19, 19, 19, …, 23 |
| OGLE-TR-132 b | 1.350 | 1.430 | 1.19 | 1.13 | 0.0298 | 0.0306 | 0.0314 | 0.001 | 0.01 | 0.03 | 1.25[a] | 20, 20, 20, 20, 20, …, 6 |
| HD 189733 b | 0.820 | 0.760 | 1.15 | 1.15 | 0.0309 | 0.0313 | 0.0317 | 0.001 | 0.01 | 0.03 | 5.26 | 21, 21, 21, 21, 22, …,23 |
| GJ 876 d | 0.320 | 0.360 | 0.018 | 0.143 | 0.0208 | 0.0208 | 0.0208 | 0.001 | 0.01 | 0.03 | 9.9 | 24, 24, 24, 25, 24, …, 7 |
| Gl 581 c | 0.310 | 0.380 | 0.016 | 0.138 | 0.0714 | 0.073 | 0.0746 | 0.09 | 0.16 | 0.23 | 2.0[b] | 26, 26, 27, 25, 27, 27, 27 |
| HD 69830 b | 0.860 | 0.895 | 0.032 | 0.167 | 0.0785 | 0.0785 | 0.0786 | 0.06 | 0.1 | 0.14 | 7.0[a] | 28, 28, 28, 25, 28, 28, 28 |

Table 1 – Physical and Orbital Parameters

References. - (1) Butler *et al.* (2006); (2) Knutson *et al.* (2007); (3) Laughlin *et al.* (2005); (4) Takeda *et al.* (2007); (5) Bakos *et al.* (2007a); (6) Gillon *et al.* (2007); (7) Saffe *et al.* (2006); (8) Winn *et al.* (2007b); (9) Melo *et al.* (2006); (10) Bakos *et al.* (2007b); (11) Sato *et al.* (2005); (12) Donovan *et al.* (2006); (13) Charbonneau *et al.* (2007); (14) Collier Cameron *et al.* (2007); (15) Pont *et al.* (2007); (16) Holman et al (2006); (17) McCollough *et al.* (2006); (18) Winn *et al.* (2007a); (19) Gillon *et al.* (2006); (20) Moutou *et al.* 2004; (21) Bakos *et al.* (2006); (22) Bouchy *et al.* (2005); (23) Burrows *et al.* (2007); (24) Rivera *et al.* (2005); (25) Sotin *et al.* (2007); (26) Bonfils *et al.* (2005); (27) Udry *et al.* (2007); (28) Lovis *et al.* (2006).

Notes: a – This value is the average of the minimum and maximum ages reported. b – This value is the minimum age reported.

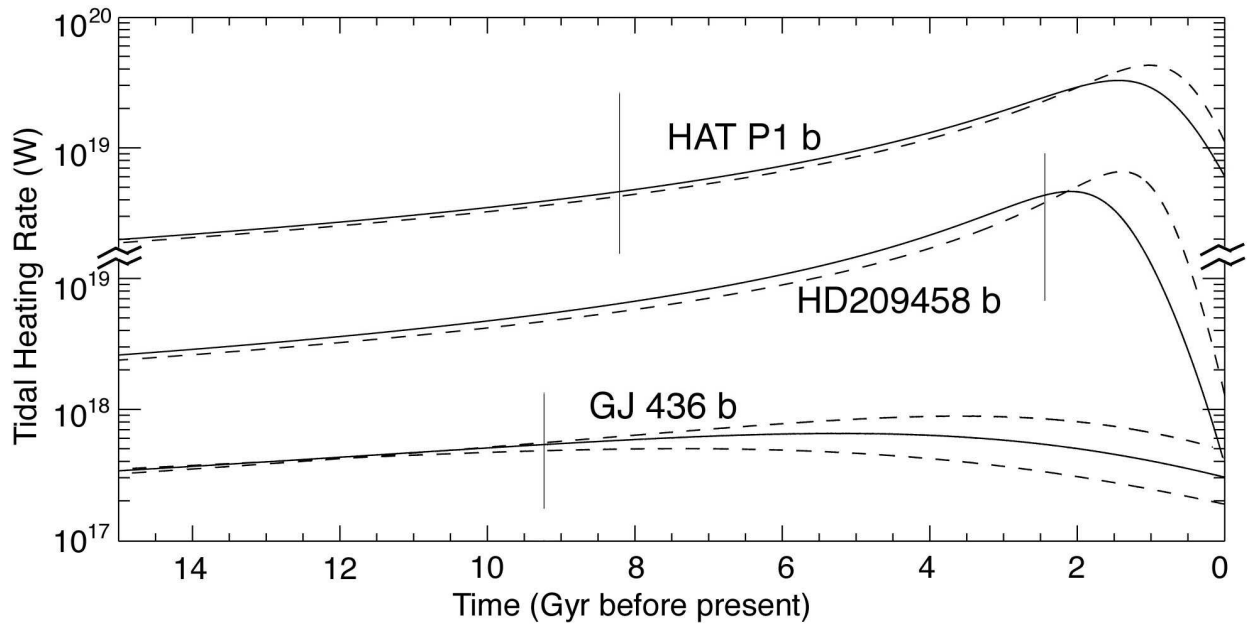

Figure 1: The tidal heating rates for planets HAT-P-1 b, HD 209458 b, and GJ 436 b as a function of time. The present time (t=0) is at the right, and the scale indicates the time before the present. Vertical lines indicate the best estimate of the formation time for the planet. Note that the vertical scale has been shifted (by a factor of 10) for HAT-P-1 b to make its curves more visible. For each planet, here and in subsequent figures, the vertical scale that corresponds to each curve is the scale intersected by that curve. The solid curve for each planet is based on the current nominal eccentricity value, while the dashed lines assume the maximum and minimum heating consistent with observational uncertainty in the orbital elements. For HAT-P-1 b and HD 209458 b, observations could not exclude a current eccentricity of zero, so the lower bound on heating rates is formally zero. Hence in those cases only one dashed line is shown, representing the upper limit. The following figures use the same conventions.

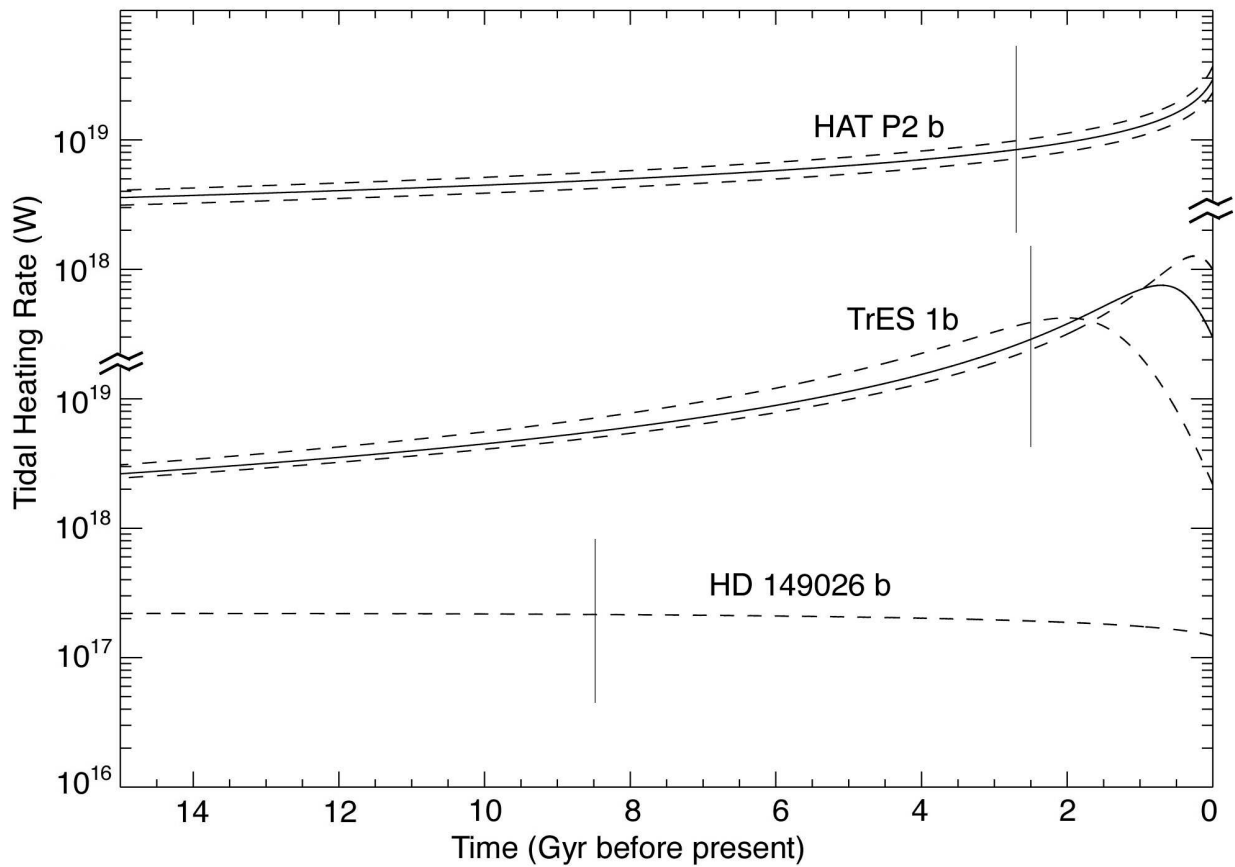

Figure 2: The tidal heating rates for planets HAT-P-2 b, TrES 1 b, and HD 149026 b as a function of time, similar to Fig. 1. Note that the vertical scale has been shifted (by a factor of 100) for HAT-P-2 b to make its curves more visible. The solid lines and dashed lines have the same meaning as in Fig. 1. For HD 149026 b, the dashed line represents the maximum heating consistent with observational limits, while the nominal eccentricity is zero, corresponding to zero heat (off scale) throughout the history.

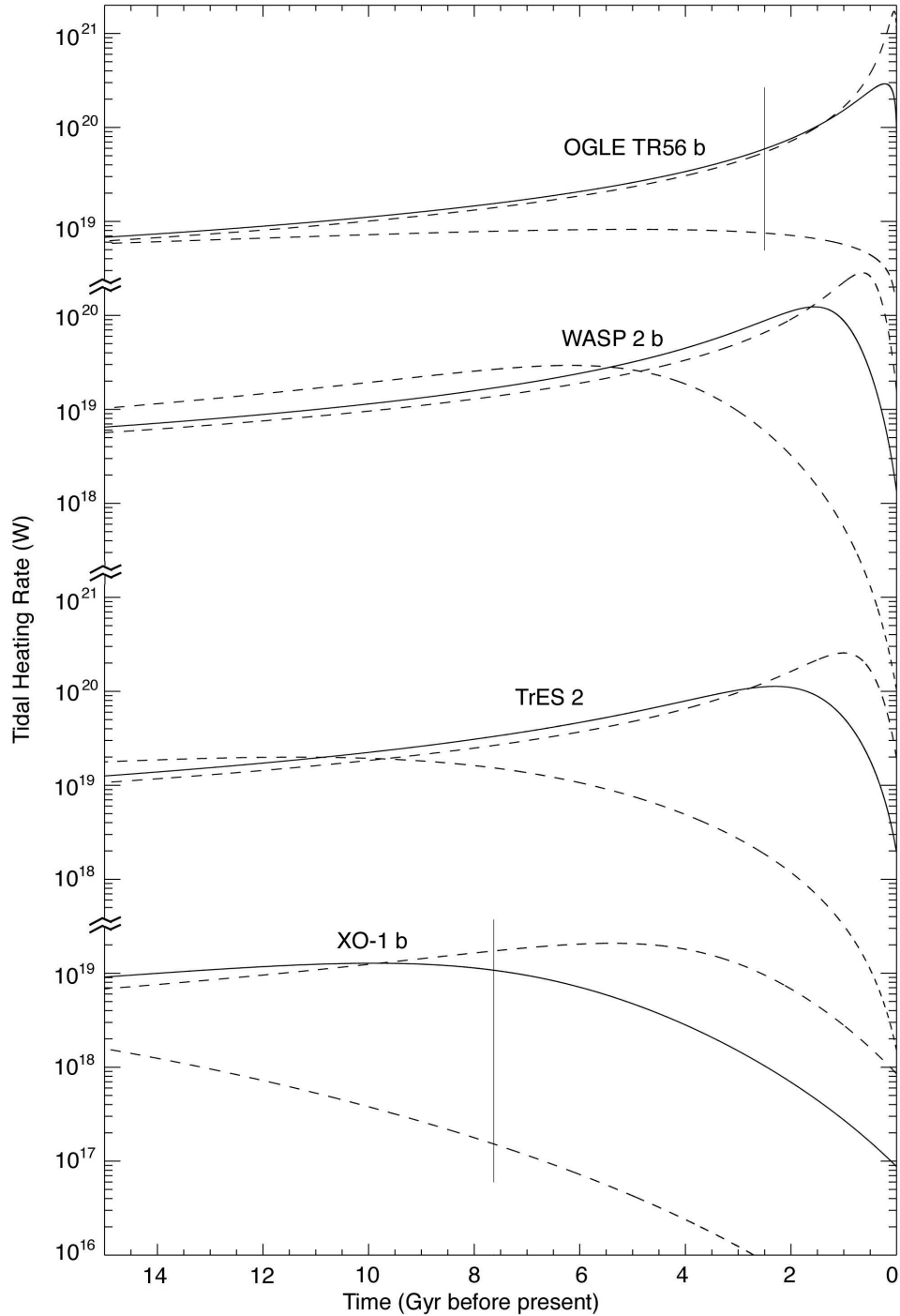

Figure 3: The tidal heating rates for four planets for which the nominal (tabulated) current *e* value is zero, but non-circular motion is likely. For each planet, a heating curve is computed for an assumed current *e* of 0.001, 0.01, and 0.03. The solid line is for 0.01, and greater current *e* corresponds to greater current heating. The scale on the ordinate is shifted by factors of ten to allow the separate results to be displayed clearly. For planets where the vertical age bar is missing, no age estimate is available.

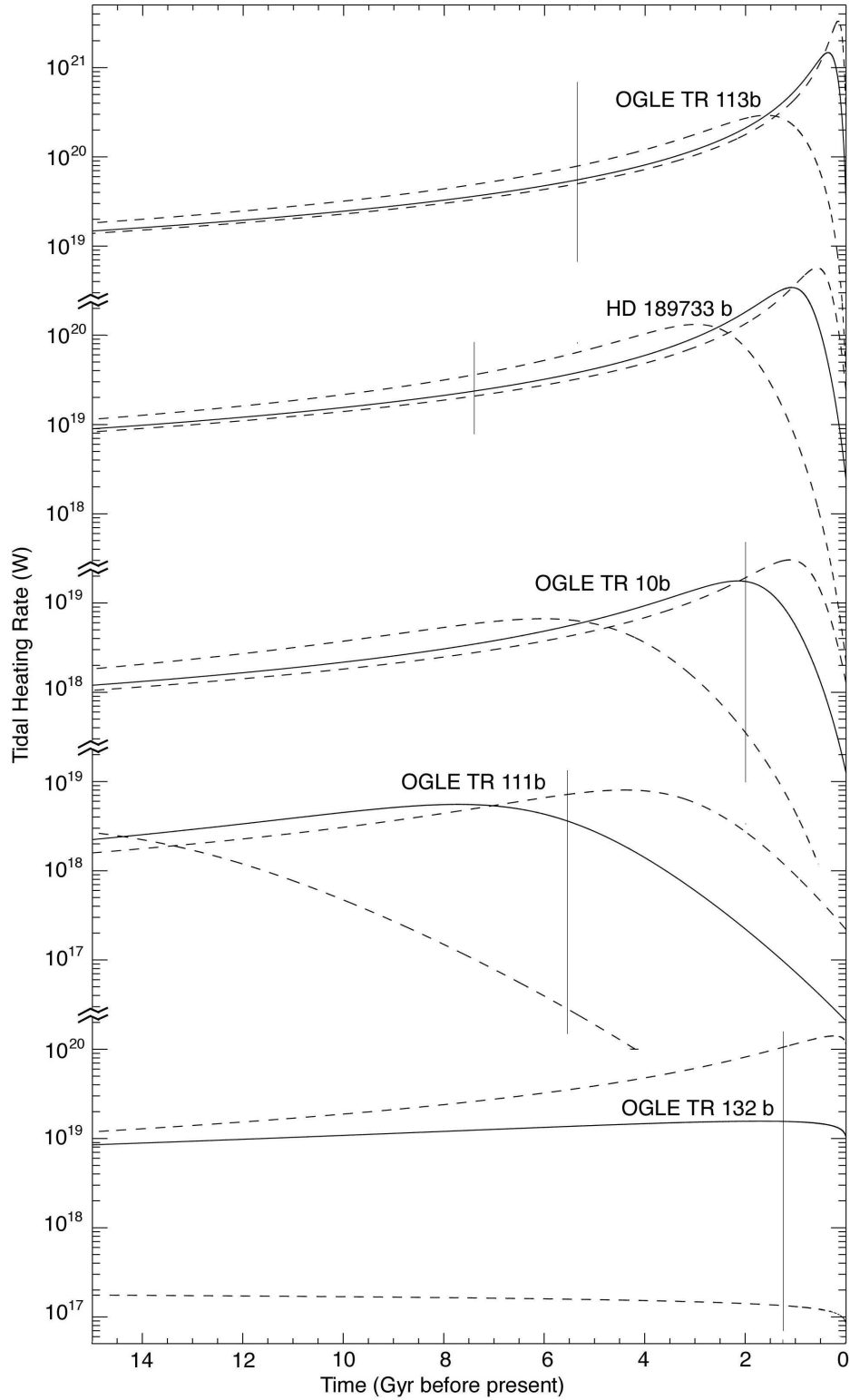

Figure 4: The tidal heating rates for five more planets for which the nominal (tabulated) current *e* value is zero, but non-circular motion is likely (similar to Fig. 3).

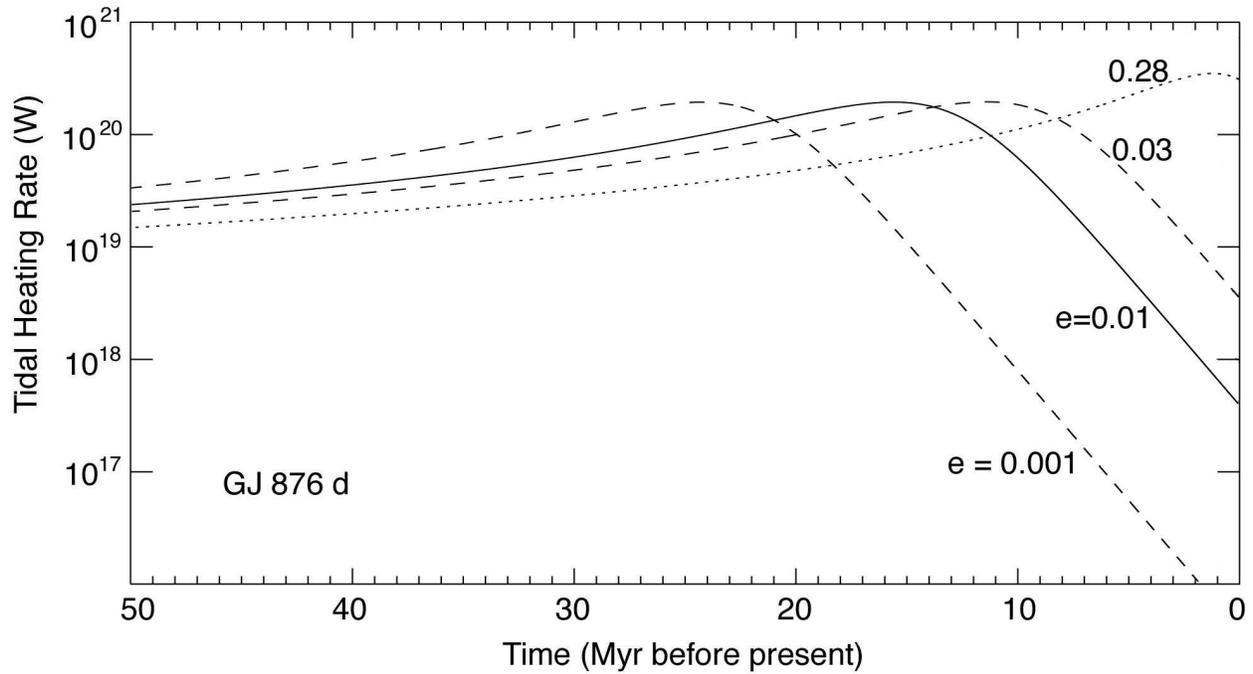

Figure 5: The tidal heating rates for GJ 876 d, a possibly terrestrial-scale for which the nominal (tabulated) current *e* value is zero, but non-circular motion is likely. A heating curve is computed for an assumed current *e* of 0.001, 0.01, and 0.03, as well as for a possible maximum value of 0.28. Extensive internal melting might be expected if the heating rate were greater than ~$10^{18}$ W. tidal heating may have been a significant factor in the geophysical history of this planet.